\documentclass[preprint]{epl}
\usepackage{graphics}
\title{Power-law resistivity behavior in 2D superconductors across the magnetic field-tuned superconductor-insulator transition}
\shorttitle{Power-law resistivity behavior}
\author{G. Sambandamurthy\inst{1,2}\thanks{E-mail: \email{gsmurthy@nhmfl.gov}} \and A. Johansson\inst{1} \and E. Peled\inst{1} \and D. Shahar\inst{1} \and P. G. Bj\"ornsson\inst{3} \and K. A. Moler\inst{3}}
\shortauthor{G. Sambandamurthy \etal}
 
\institute{
  \inst{1} Department of Condensed Matter Physics, Weizmann Institute of Science - Rehovot 76100, Israel\\
  \inst{2} National High Magnetic Field Laboratory - Tallahassee, FL 32306, USA\\
  \inst{3} Department of Applied Physics, Stanford University - Stanford, CA 94305, USA.
}
\pacs{74.25.Fy}{Transport properties}
\pacs{74.78.-w}{Superconducting films and low-dimensional structures}
\pacs{74.25.Dw}{Superconductivity phase diagrams}
\begin{document}

\maketitle

\begin{abstract}
We present the results of a systematic study of thin-films of amorphous indium-oxide near the superconductor-insulator transition. We show that the film's resistivity follows a simple, well-defined, power-law dependence on the perpendicular magnetic field. This dependence holds well into the insulating state. Our results suggest that a single mechanism governs the transport of our films in the superconducting as well as insulating phases.
\end{abstract}

At temperatures ($T$s) near the absolute zero, the superconductor-insulator transition (SIT) in two-dimensional systems is a dramatic phenomenon. Over a rather narrow stretch of parameters, such as magnetic field ($B$) or film thickness, the resistivity ($\rho$) swings from being immeasurably low, essentially zero, to being exponentially diverging with lowering $T$ \cite{Goldman98}. One does not expect, given this large disparity in the behavior of $\rho$, that a unified description of transport in these two opposing regimes should exist.

It is therefore surprising that a theoretical framework was developed, in which this common description naturally emerges \cite{Fisher90,Fisher91}. Since the insulator and the superconductor are two distinct $T=0$ phases of the electronic system, the SIT is considered as a quantum phase transition (QPT), driven by a parameter in the Hamiltonian that can, in principle, be controlled in experiments \cite{Sondhi97}. Within this framework the resistivity, in both the superconducting and insulating phases, is described by a single universal scaling function that is expected to be relevant in the vicinity of the transition. Evidence for, and against, the validity of the QPT approach to real samples has been reported in the literature \cite{Hebard90,Liu91,Yazdani95,Gantmakher00_2,Mason01,Kapitulnik01,Wu02}.

The purpose of this Letter is to show that the resistivity of our superconducting amorphous indium-oxide (a:InO) films can be described by a single function covering a wide range of our measurements, which includes the $B$-driven SIT. This function can be written as follows:
\begin{equation}
\rho(B,T)=\rho_{c}(\frac{B}{B_c})^{T_{0}/2T}
\label{rlaw}
\end{equation}
where $\rho_{c}$, $B_c$ and $T_{0}$ are sample-specific parameters.  

The phenomenological form introduced above is consistent with the collective-pinning model of transport in thin superconducting films, which predicts a vortex-pinning energy proportional to $\ln(B)$ \cite{Blatter94a}. This form has been observed before in high-$T_c$ layered systems \cite{Palstra88,Inui89} as well as in amorphous superconductors \cite{White93,Chervenak96,Ephron96}. The new result of our work is that this behavior is not restricted to the superconducting phase but continues, uninterrupted, well-into the insulating state \cite{Shahar98}. 

Our data were obtained from a detailed study of disordered thin-films of a:InO. The films were prepared by e-gun evaporating high purity (99.999 \%) \chem{In_2O_3} on clean glass substrates in a high vacuum system. The thickness was measured {\it in-situ} by a quartz crystal thickness monitor. Lithographic techniques were used to pattern the films to Hall-bars with voltage probe separation twice the with of the Hall bar. Data from 30 different samples with widths ranging from 2 $\mu$m to 500 $\mu$m are reported in this study. We tuned the inherent disorder in the samples, and hence its low $T$ behavior, by thermal annealing of the films as described in \cite{Murthy04, Murthy05}. Resistance measurements were carried out in the four-probe configuration by low frequency AC lock-in techniques, with excitation currents of 10 pA--10 nA. The samples were cooled either in a dilution refrigerator with base $T$ of 0.01 K or in a He-3 refrigerator with base $T$ of 0.25 K. 

\begin{figure}
\includegraphics[width=4in]{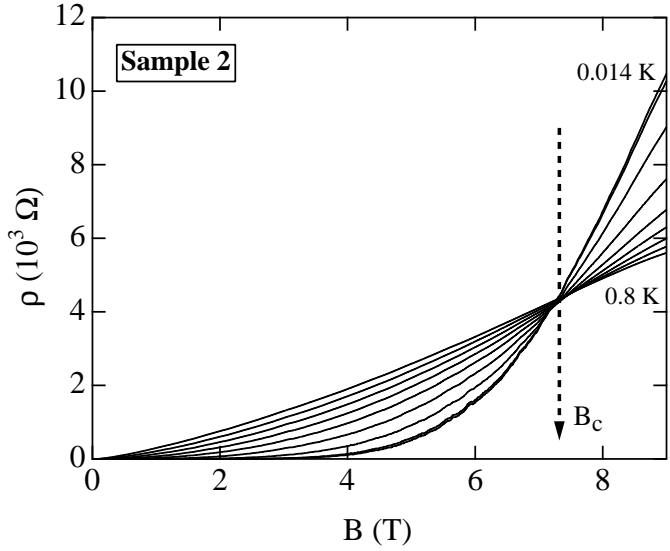}
\caption{Sheet resistance ($\rho$) as a function of $B$, of one of our samples, measured at $T=$ 0.014 K, 0.1 K, 0.2 K, 0.3 K, 0.4 K, 0.6 K, 0.7 K and 0.8 K. The vertical arrow marks the crossing point of different $\rho$ isotherms, that identifies $B_c$. $B_c$ = 7.31 T for this sample.}
\label{R_B_T}
\end{figure}

In Figure \ref{R_B_T} we show $\rho$ vs. $B$ taken at several $T$s below $T_c$ which, for this sample, was $1.5$ K. A clear and well-defined crossing of the various $\rho$ isotherms is evident at $B=7.31$ T. This point, termed $B_c$,  has been traditionally associated with the SIT. This is a natural viewpoint, for four reasons. First, since the determination of the phase of the system is done by extrapolating the $T$-dependence $\rho$ data to $T=0$, a $B$ value where the temperature coefficient of resistivity changes sign at low $T$s is taken to indicate the phase transition point. The existence of a sharp and well-defined crossing point, which is the second reason, is in accordance with theoretical predictions. Third, also in agreement with  theory, scaling behavior near the crossing $B$-point, observed over a limited $T$ range, has been reported by several groups \cite{Liu91,Gantmakher00_1,Gantmakher00_2,Mason01}.

\begin{figure}
\includegraphics[width=4in]{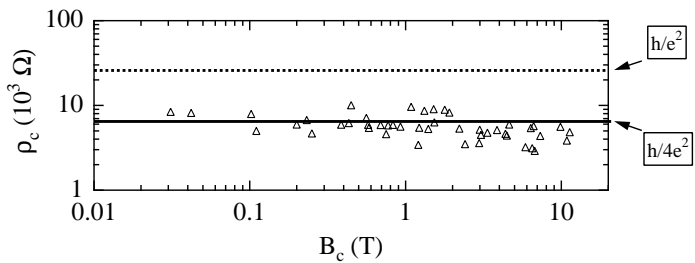}
\caption{Critical resistance at the crossing point for our superconducting samples (thickness between 20 and 30 nm) are plotted against the corresponding $B_c$ values. The solid horizontal line marks $h/4e^2$ and the dashed horizontal line marks $h/e^2$, respectively. $\rho_c$ values are scattered around 5.8 $k\Omega$, with a standard deviation of 1.8 $k\Omega$.}
\label{Rc_Bc}
\end{figure}

The fourth reason can be seen in Figure \ref{Rc_Bc}, where we plot the $\rho$ value at $B_c$, $\rho_c$, obtained from all our superconducting samples that exhibited a well-defined crossing point. The data are scattered around $5.8$ k$\Omega$ and, with a standard deviation of $1.8$ k$\Omega$.
A scatter of about a factor of 3 in $\rho_{c}$ can not usually be taken as an indication of a universal number. However, if we consider the fact that the measured $\rho$ itself spans more than 10 orders of magnitude in value, a three-fold variation does not seem that large, indicating that the Cooper-pair quantum resistance $h/4e^{2}$ ($\approx$6.45 k$\Omega$) is of special significance for the transition.
This value is in good agreement with the theoretical prediction for this transition \cite{Fisher90}, and with other experimental studies in the literature \cite{Hebard90,Liu91}, with the notable exception of experiments done on MoGe \cite{Yazdani95}. Taken together these points present a compelling case for the validity of the QPT approach and to the identification of $B_c$ with the transition point.

A central assumption that underlies the QPT approach to the $B$-driven SIT problem is that, on a microscopic level, the nature of the transport process is not significantly altered as one crosses the transition $B$ into the insulating phase. In other words this means that, locally, superconductivity must persist beyond the transition. According to Fisher \cite{Fisher90}, the transition is manifested by a change in the macroscopic vortex-state, and Cooper pairs must still exist, albeit localized, in the insulating phase to support the formation of vortices. While previous experiments designed to test this assumption resulted in conflicting conclusions \cite{Valles92,Hsu95,Markovic99}, we will argue below that our results are in strong support of its validity.

We begin by taking, for the moment, the standpoint in which the transition to the insulator coincides with the complete disappearance of superconductivity in our type-II films, i.e., $B_c=H_{c2}$. We next show that this standpoint leads to a conflict with the experimental results, requiring a nonphysically large variation in the value of $\xi$, the superconducting coherence length. 

Consider the data presented in Figure \ref{Rc_Bc} that, aside from being consistent with a universal value of $\rho_c$, have the following implication. These data were obtained from 30 different samples at 46 different anneal stages, spanning a range of disorder that, although hard to quantify, can be specified by the normal-state resistivity of the samples, $R_N$. For our superconducting samples, $R_N$ and therefore the mean free path $l$ change by no more than 50\%. Through the relation $\xi=\sqrt{\xi_{0}l}$ ($\xi_{0}$ is the superconducting coherence length in the clean limit), we conclude that the variation in $\xi$ are limited to less than 50\%. The contradiction with the $B_c=H_{c2}$ assumption arises when we recall that $H_{c2}=\frac{\Phi_0}{2\pi\xi^2}\propto l$, which clearly can not account for more than two orders of magnitude variation in the observed $B_c$. We therefore conclude that the crossing point at $B_c$ can be at much lower field than the superconducting critical field $H_{c2}$. 

\begin{figure}
\includegraphics[width=4in]{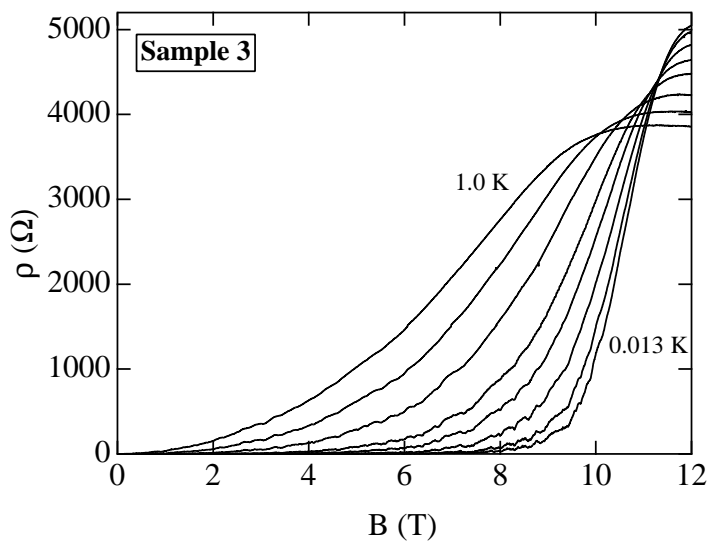}
\caption{$\rho$ of a low-disordered sample as a function of $B$ at $T=$ 0.013 K, 0.1 K, 0.2 K, 0.3 K, 0.4 K, 0.6 K, 0.8 K and 1 K. Distinct crossing point of the different isotherms is absent and the transition is smeared over 1.5 T, with the center around 11.32 T.}
\label{noXing}
\end{figure}

This brings up the question of the identifications of $H_{c2}$ in high-disorder, thin-film superconductors \cite{Smith00}. In Figure \ref{noXing} we present $\rho$ vs. $B$ at several $T$s obtained from a lower disorder sample. Two features are notable in this graph. First, superconductivity survives to a large $B$, around 11 T, and second, the crossing point of the $\rho$ isotherms is clearly not present, the transition being smeared over approximately 1.5 T. This smearing is expected for the thermodynamic $H_{c2}$ which should depend on $T$. We also note that the critical $B$ seem to have a limiting value in our a:InO samples of around 12 T. In a recently published Letter \cite{Murthy04}, we provided evidence to the existence of a relation between the superconductor $T_c$ and $T_I$, the temperature which characterize the transport in the $B$-induced insulating state terminating the superconducting phase. We found that the $B$ position of the insulating peak is only weakly dependent on disorder and appears in the range of 8-12 T. A possibility therefore exists that the true $H_{c2}$ of our films is near this value. 

\begin{figure}
\includegraphics[width=4in]{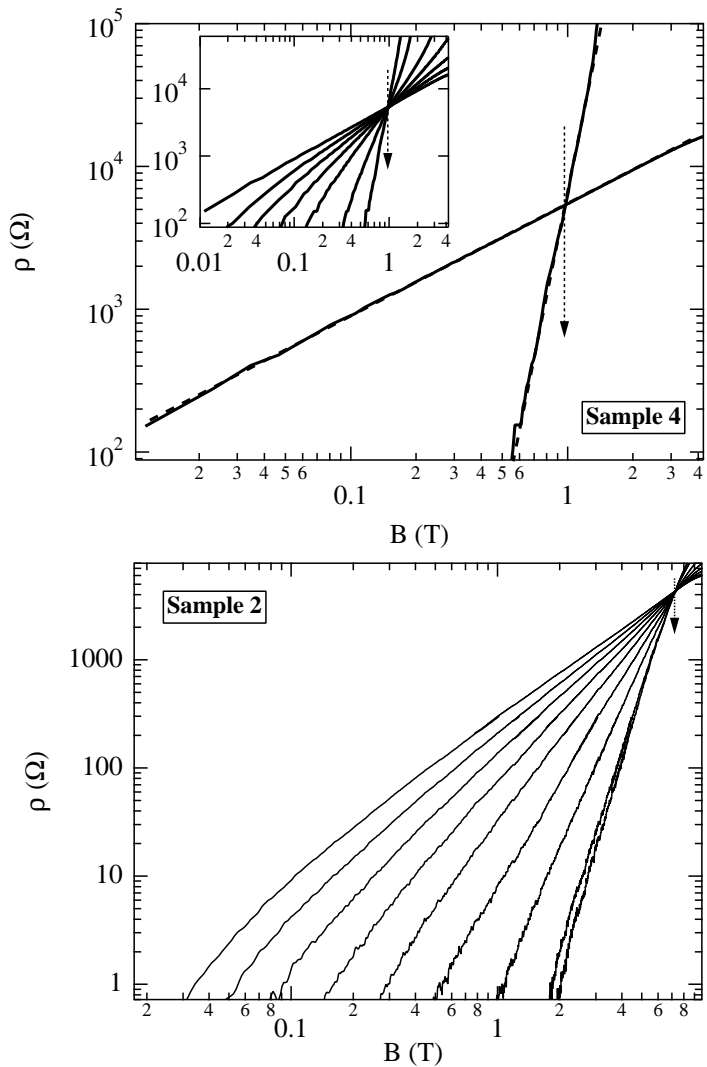}
\caption{$\rho$ versus $B$ isotherms for two of our samples presented in log-log graphs. Upper panel : Isotherms at $T$= 0.02 and 1.2 K are only shown for clarity. The inset shows the full set taken at $T$= 0.02 K, 0.2 K, 0.4 K, 0.6 K, 0.8 K,1.0 K and 1.2 K. Lower panel : Data from another sample taken at same $T$ values as in Fig. 1. Vertical arrows mark the crossing point of different $\rho$ isotherms. The $\rho$ isotherms display a power-law dependence on $B$, shown by the dotted lines in the main figure on the upper panel, that holds on either side of the crossing point.}
\label{powerLaw}
\end{figure}

\begin{figure}
\includegraphics[width=4in]{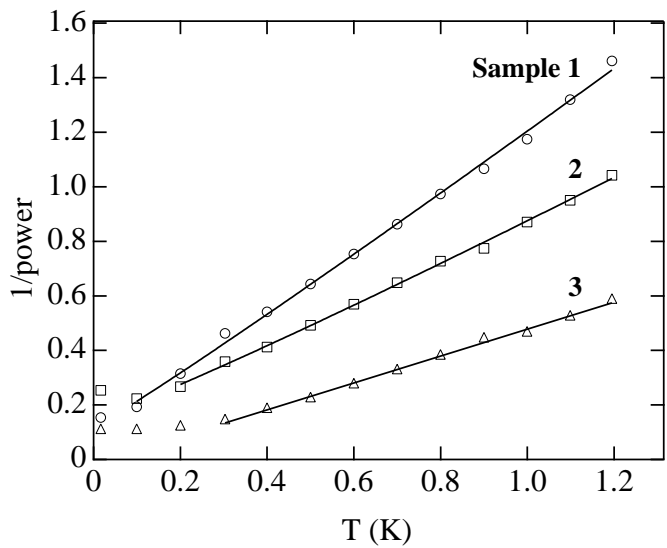}
\caption{The inverse of the power extracted from the power-law dependence of the $\rho$-$B$ isotherms as a function of $T$ for 3 different samples. The solid lines are straight line fits. The $T_0$ values obtained from the fits are 1.8 K, 2.5 K and 4 K for the 3 samples. The sample parameters are listed in Table I. Deviations from the linear fits are observed at low-$T$ for all the samples.}
\label{powers}
\end{figure}

We next perform a quantitative analysis of the $B$ and $T$-dependence of our resistivity data, which will lead us to the central result of our work. In Figure \ref{powerLaw} we again plot $\rho$ vs. $B$ at various $T$'s for two of our samples, but this time we use log-log graphs. For the sample in the upper panel of Figure 4 we took special care to extend our measurements over a large range of $\rho$. Each curve is well-described by a power-law dependence that holds over more than 2 orders of magnitude in $B$ and more than 3 in $\rho$, with non-random deviations that are only seen at high $T$s and as the $B$ values approach the insulating peak. The different curves are distinguished by their power, which is a function of $T$. Our entire data can be summarized by the following expression:
\begin{equation}
\rho(B,T)=\rho_{c}(\frac{B}{B_c})^{P(T)}.
\label{rlaw2}
\end{equation}
To delineate the form of $P(T)$ it is convenient to plot its inverse versus $T$, see Figure \ref{powers}. A linear description best fits the data, with the parameter $T_0$ being close to twice the value of $T_c$ of the film at $B=0$. The final form is presented in Eq.\ref{rlaw}. The parameters for the three samples shown in Figure \ref{powers} are presented in Table I. Recent experiments \cite{Gantmakher01,Baturina04} have used the $T$ at which the power $P(T)$=1 to determine the $T_c$ at $B$=0. Following that description, in our data in Figure \ref{powers}, for sample 1, $P(T)$=1 at $T$=0.82 K, as compared to the $T_c$=1 K obtained from our $\rho - T$ curves.

This leads us to a discussion on the origin of the behavior of Eq.\ref{rlaw}, providing the theoretical basis for the experimental observation that $T_{0}/2=T_{c}$. A power-law $B$-dependence of $\rho$ in two-dimensional superconductors is associated with the collective-pinning flux-creep transport model, predicting an activation (pinning) energy, $U_0$, that depends logarithmically on $B$ \cite{Blatter94a,Palstra88,Inui89,White93,Chervenak96,Ephron96}: 
\begin{equation}
U_0=\frac{1}{2}T_{c}ln(B_c/B).
\label{pinE}
\end{equation}
This, in association with activated transport that can be described by 
\begin{equation}
\rho(B,T)=\rho_{c}e^{{-T_0(B)}/{T}},
\end{equation}
leads to the power-law dependence of Eq.\ref{rlaw2}. Here we use $T_0$ as the sample-specific activation energy extracted from our data and $U_0$ as the activation energy as described in the collective pinning theory, though they imply the same physical meaning. Similar power-law behavior in $\rho$ was observed in disordered thin-films \cite{White93,Chervenak96,Ephron96} and in layered high-$T_c$ compounds \cite{Palstra88,Inui89}, and may be indicative of the central role played by vortices in our system. 

Reiterating the intriguing feature in our results we note that the power-law behavior described by Eq.\ref{rlaw2} continues, uninterrupted, through $B_c$ and into the insulating state (see inset of Fig.\ref{powerLaw}a). Though in Eq.~\ref{rlaw}, $B_c$ refers to the theoreticall upper critical field value, experiments on disordered thin-films have found \cite{Chervenak96} the activation energy to go to zero for $B$ values much lower than the upper critical field. Inspecting Figure \ref{powers} reveals another aspect of the data related to the limiting low-$T$ behavior. Below $0.2$ K, $1/P(T)$ deviates from its high-$T$ linear dependence and seems to saturate. This directly implies a saturation of $\rho(B,T)$ at low-$T$, similar to that observed by Ephron {\it et al.} in MoGe \cite{Ephron96}. At present we are unable to ascertain the reason for this saturation in our samples.

To conclude, although our observation that the behavior represented by Eq.\ref{rlaw} straddles both sides of $B_c$ lends support to the validity of the central assumption of the QPT approach, it does not constitute a verification of the QPT approach as it applies to the SIT. In order to do that it is necessary to first clarify the nature of the apparent low-$T$ saturation of the resistivity near the transition. Additionally, a detailed experimental work is needed to establish the existence and nature of the divergence at the transition. Only then it will be possible to make a significant evaluation of the validity of the QPT theory to our samples.

We wish to thank Z. Ovadyahu, Y. Oreg, S. L. Sondhi and D. Huse for useful discussions. This work is supported by the ISF, the Koshland Fund and the Minerva Foundation. P.G.B. and K.A.M. acknowledge the support from NSF grant DMR-9802719. 

\begin{table}
\caption{Parameters for the 3 samples in Fig. 5. $R_N$ values are measured at $T$ = 4.2 K}
\begin{center}
\begin{tabular}{|c|c|r|r|c|c|}
\hline
Sample&$d$ (nm)&$R_N$ (k$\Omega$)&$B_c$ (T)&$T_c$ (K)&$T_0$ (K)\\
\hline
1&30&4.9&2.41&1.0&1.8\\
2&30&3.47&7.31&1.5&2.5\\
3&30&3.11&11.32&1.9&4.0\\
\hline
\end{tabular}
\end{center}
\end{table}

\end{document}